\documentclass[final, times, twocolumn, 2p]{elsarticle}

\usepackage{tikz}
\usepackage[ruled,vlined]{algorithm2e}
\usepackage{graphicx}
\usepackage{dcolumn}
\usepackage{bm}
\usepackage{amsmath,amssymb}    
\usepackage{url}
\newcommand{\beq}{\begin{equation}}
\newcommand{\eeq}{\end{equation}}

\journal{}
\begin{document}
\begin{frontmatter}

\title{On the performance limits of coatings for gravitational wave detectors made  of alternating layers of two materials}

\author[unisannio,INFN]{V. Pierro}
\author[unibas,INFN]{V. Fiumara}
\author[unisaING,INFN]{F. Chiadini}
\author[unisaFIS,INFN]{F. Bobba}
\author[unisaFIS,INFN]{G. Carapella}
\author[unisaFIS,INFN]{C. Di Giorgio}
\author[unisaFIS,INFN]{O. Durante}
\author[CNR,INFN]{R. Fittipaldi}
\author[unisannio,INFN]{E. Mejuto Villa}
\author[unisannio,INFN]{J. Neilson}
\author[unisaFIS,MUSEOEF,INFN]{M. Principe}
\author[unisannio,MUSEOEF,INFN]{I. M. Pinto}

\address[unisannio]{Dipartimento di Ingegneria,  Universit\'{a} del Sannio, I-82100 Benevento, Italy.}
\address[unibas]{Scuola di Ingegneria, Universit\'{a} della Basilicata, I-85100 Potenza, Italy.}
\address[unisaING]{Dipartimento di Ingegneria Industriale DIIN, Universit\'{a} di Salerno, I-84084 Fisciano, Salerno, Italy.}
\address[unisaFIS]{Dipartimento di Fisica "E.R. Caianiello", Universit\'{a} di Salerno, I-84084 Fisciano, Salerno, Italy.}
\address[CNR]{CNR-SPIN, c/o Universit\'{a} di Salerno, I-84084 Fisciano, Salerno, Italy.}
\address[MUSEOEF]{Museo Storico della Fisica e Centro Studi e Ricerche "Enrico Fermi", I-00184  Roma, Italy.}
\address[INFN]{INFN, Sezione di Napoli Gruppo Collegato di Salerno,  Complesso Universitario di Monte S. Angelo, I-80126 Napoli, Italy.}

\begin{abstract}
The coating design for mirrors used in interferometric detectors of gravitational waves currently consists of stacks 
of two alternating dielectric materials with different refractive indexes. 
In order to explore the performance limits of such coatings, 
we have formulated and solved the design problem as a multiobjective optimization problem 
consisting in the minimization of both coating transmittance and thermal noise.
An algorithm of global optimization (Borg MOEA) has been used without any {\it a priori} assumption on the number and thicknesses of the layers in the coating. 
The algorithm yields to a Pareto tradeoff boundary exhibiting a continuous, 
decreasing and non convex (bump-like) profile, bounded from below 
by an exponential curve which can be written in explicit closed form in the transmittance-noise plane. 
The lower bound curve has the same expression of the relation between transmittance and noise for the quarter 
wavelength design where the noise coefficient of the high refractive index material assumes a smaller equivalent value.
An application of this result allowing to reduce the computational burden of the search procedure  is reported and discussed.
\end{abstract}

\begin{keyword}
Dielectric multilayers, Gravitational Wave Interferometric Detectors, Multiobjective optimization\\
\vspace{0.1cm}
{\em DOI:}	10.1016/j.optmat.2019.109269


\end{keyword}

\end{frontmatter}

\section{Introduction}
The first direct detection of gravitational waves (henceforth GW) by the LIGO \cite{LIGO} and Virgo \cite{Virgo} detectors, and the recent Multi Messenger
observation of GW170817 - GRB 170817A - SSS17a/AT 2017gfo marked the birth of Multi Messenger Astronomy (MMA) \cite{MMA1,MMA}.
Increasing the visibility distance of the operating GW detectors is a needed step to deploy the full potential of MMA.
Thermal (Brownian) fluctuations in the high-reflectance (HR) coatings of the test-masses is presently the dominant noise source in
interferometric GW detectors \cite{Abernathy18} setting their ultimate visibility distance in the (40-300) Hz band, 
where all recent detections have been made.  
Notably, efforts to reach and beat the quantum noise limit will be
meaningful only after a significant reduction of coating thermal noise is achieved.

Reducing coating thermal noise is thus the top current priority in GW detectors R\&D \cite{Flaminio10}.
The current research direction in coating technology explore two promising options for reducing coating thermal noise as needed by
new generation detectors, namely the search of new material (e.g. crystalline materials \cite{cryst} nm-layered composite materials \cite{nanolayer}, 
Silicon Nitrides \cite{SiliconNitride1,SiliconNitride} etc.), and the  optimization of coating design, and deposition parameters to achieve
the best relevant figures of merit (low optical and mechanical losses, high optical contrast).

In this paper we focus on optimization of the coating structure adopted in advanced LIGO and advanced Virgo. 
High reflectance test-mass coatings consist of multilayers of alternating low and high refractive index materials 
(silica and 14.5\% titania-doped tantala in the advanced LIGO and Virgo \cite{pinard11, pinard17}),
which must provide the required reflectance with minimal thermal noise  \cite{Harry}.
 
High reflectance  optical coatings typically consist  of a stack of identical high/low index layer pairs or {\em doublets,}
where each layer is quarter wave thick at the operating frequency \cite{pinard11}.
This design features the minimum number of layers to achieve some specified reflectance,
but does not yield the minimum noise among all possible iso-reflective designs (see \cite{Harry} chapter 12).

An alternative design consisting of a stack of identical doublets (with the exception of the terminal top/bottom doublets) 
with non-quarter wave layers was proposed in \cite{Spie, Villar10} and further explored in \cite{Russian}, 
and shown to outperform the classical quarter wavelength design.



In this paper, 
we formulate for the first time to the best of our knowledge a multiobjective coating optimization problem 
consisting in the minimization of both coating transmittance and 
thermal noise. We use a global optimization method (Borg MOEA \cite{borgMOEA}), making no {\it a priori}
assumptions about layer thicknesses, and  we find a simple closed form lower bound of the general coating performance in the
transmittance-noise plane.
Also, the end-tweaked stacked-doublet coating structure
of the {\it optimized} coatings 
assumed in \cite{Harry,Spie,Villar10} on the basis of
partial evidences, is obtained in a rigorous way.

In the following an $\exp( \imath \omega t)$  dependence on time $t$ is implicit, where $\omega$
is the angular frequency and $\imath$ is the imaginary unit.

The paper is organized as follows: in Sect. 2 and 3 we introduce the optical and thermal-noise model. 
Constrained optimization and the multiobjective formulation are discussed in Sect. 4. The results are collected in Sect. 5, conclusions and recommendations for 
future research follow in Sect. 6. 

\begin{figure}
\centering
\includegraphics[width=8 cm]{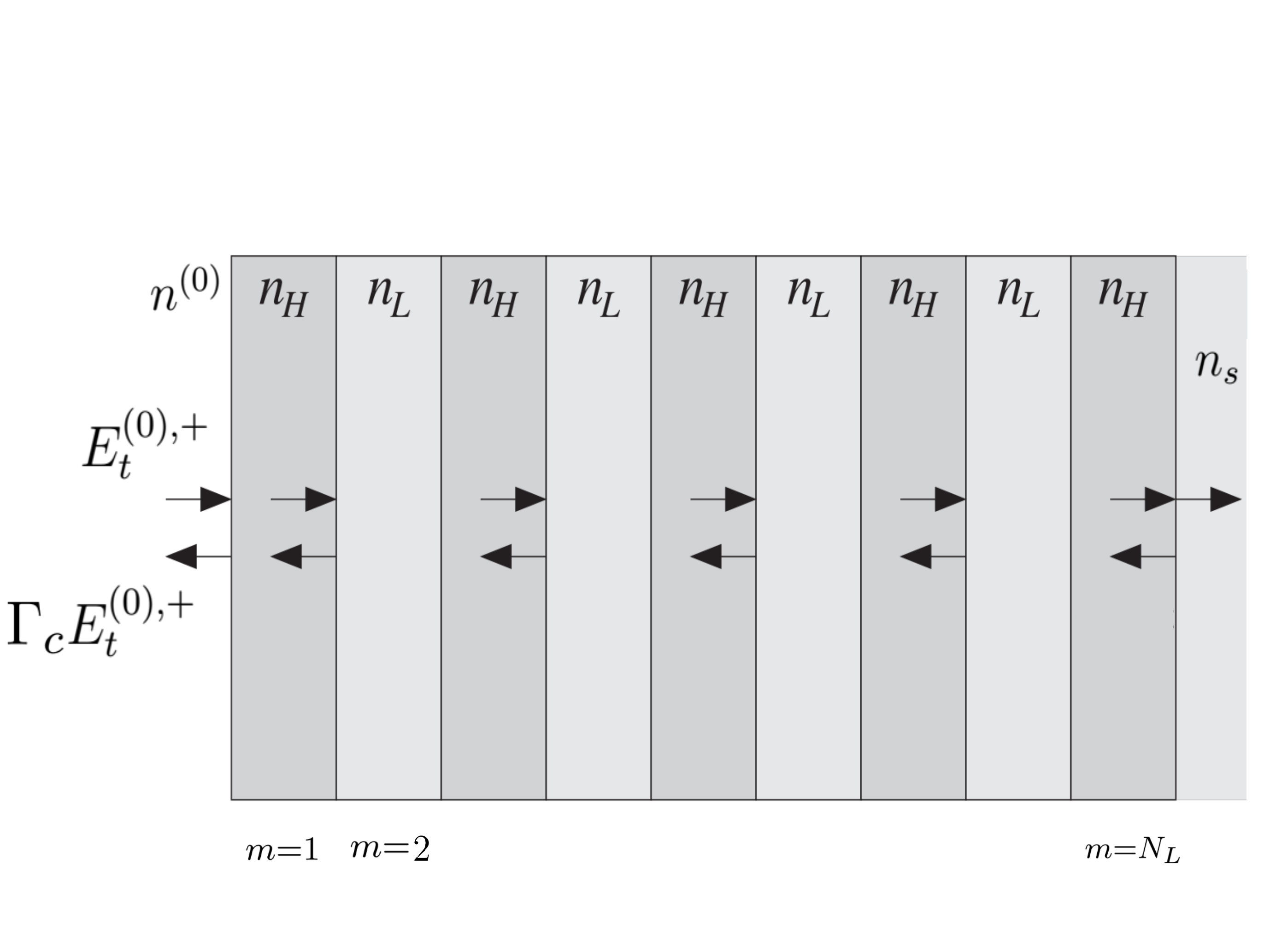}
\caption{Multilayer structure made of $N_L$ alternating high and low refractive indexes 
denoted as $n_H$ and $n_L$ respectively, deposited on a substrate of refractive index $n_s$.
The layer index $m$ increases form left to right, the leftmost vacuum half-space corresponds
to $m=0$. The field $E_t^{(0),+}$ is the complex amplitude of the incident 
electric field (in the frequency domain).
}\label{fig:1}
\end{figure}

\section{Coating optical modeling}
Let us consider a coating consisting of a multilayer placed between two homogeneous dielectric half-spaces
with refractive indexes $n^{(0)}$ and $n_s$, respectively (see Fig. 1).
The rightmost half-space (with refractive index $n_s$) is the substrate, while the leftmost one is the vacuum.
Let a monochromatic plane wave impinge normally  on the coating from the vacuum. 
The optical reflections and transmission properties of a multilayer structure can be
computed in a closed form using the characteristic matrix method \cite{Abeles, BornWolf}.  

The characteristic matrix of the $m-$ th layer can be written \cite{Orfanidis}:
\begin{equation}
\mathbf{T}_m=
\begin{bmatrix}\tabcolsep6pt
\cos\left(\psi_m \right)&  \imath \displaystyle  (n^{(m)})^{-1} \sin\left(\psi_m\right)\\[10pt]
\displaystyle \imath n^{(m)} \sin\left(\psi_m\right)&
\cos\left(\psi_m\right)
\end{bmatrix},
\label{Tmatrix}
\end{equation}
where 
\beq
\psi_m=\frac{2 \pi}{\lambda_0}n^{(m)}d_m,
\label{optwidth}
\eeq
$\lambda_0$ and $d_m$ are the free space wavelength and the layer thickness, respectively, and $n^{(m)}$ is the complex refractive index
\beq
n^{(m)} = n^{(m)}_r - \imath \kappa^{(m)}
\eeq
where $\kappa^{(m)}$ is the extinction coefficient\footnote{ In view of the assumed $\exp(\imath \omega t)$ dependence on the time $t$, absorbing media have $\kappa^{(m)} >0$.}.

The optical response of the whole coating can be computed from the multilayer characteristic matrix,
\beq
\mathbf{T}=\mathbf{T}_1 \cdot \mathbf{T}_2 \cdot ... \cdot \mathbf{T}_{N_L}
\label{charmat}
\eeq
where $N_L$ is the total number of layers numbered from the vacuum to the substrate as illustrated in Fig. 1.

The complex reflection coefficient $\Gamma_c$ at the vacuum/coating interface is given by: 
\beq
\Gamma_c=\frac{n^{(0)}-n_c}{n^{(0)}+n_c}
\eeq
where $n_c$ is the effective refractive index of the whole multilayer structure,
\beq
n_c=\frac{T_{21}+n_s T_{22}}{T_{11}+n_s T_{12}}.
\eeq
The power transmittance at the vacuum/coating interface is $\tau_c=1 - |\Gamma_c|^2$.

The average power density  dissipated  within the coating can be computed as the difference between the
average power density flowing into the coating at the first interface (vacuum/coating) $\mathcal{P}_{in}$ and the power density 
flowing into the substrate (last interface ) $\mathcal{P}_{out}$. 
The input power is
\beq
\mathcal{P}_{in}= n^{(0)} \frac{1}{2 Z_0}|E_t^{(0),+}|^2 \tau_c
\eeq
where $E_t^{(0),+}$ is the incident transverse electrical field at the vacuum/coating interface and 
${\displaystyle Z_0=\sqrt{ \mu_0/\epsilon_0}}$ is the characteristic impedance of the vacuum.
The power density flowing into the substrate is given by using the definition of Poynting vector as follow 
\beq
\mathcal{P}_{out}= \frac{1}{2}\Re( E_t^{(N_L)}H_t^{(N_L)*})
\eeq
where $E_t^{(N_L)}$ and $H_t^{(N_L)}$ are total transverse electric and magnetic fields at the
last interface. 
The complex amplitude $E_t^{(N_L)}$ and $Z_0 H_t^{(N_L)}$ are 
obtained from $E_t^{(0)}=E_t^{(0),+} (1+\Gamma_c)$ and
 $Z_0 H_t^{(0)}=n_T^{(0)} E_t^{(0),+} (1-\Gamma_c)$
using the formula 
\beq
\begin{bmatrix}
    E_t^{(N_L)}\\
   Z_0 H_t^{(N_L)}
\end{bmatrix}
= \mathbf{T}^{-1} 
\begin{bmatrix}
    E_t^{(0)}\\
    Z_0 H_t^{(0)}
\end{bmatrix}.
\eeq

\section{The thermal noise model}
The power spectral density of coating thermal noise is given, under suitable simplifying assumptions, by \cite{Harry}
\beq
S_{coat}^{B}=\frac{2 k_B T}{f \pi^{3/2}}\frac{1-\sigma_s^2}{w Y_s}\phi_c
\label{noise}
\eeq
where $f$ is the frequency, $T$ is the (absolute) temperature, $w$ is the(assumed Gaussian) laser-beam waist, and
$\sigma_s$ and $Y_s$  are the Poisson and Young modulus of the substrate, and
the coating loss angle $\phi_c$ is
\beq
\phi_c= \sum_{m=1}^{N_L} \eta_m d_m 
\eeq
where 
\beq
\eta_{m}= \frac{1}{\sqrt{\pi} w}  \phi_{m} \left( \frac{Y_{m}}{Y_s} + \frac{Y_s}{Y_{m}} \right)
\eeq
$\phi_{m}$ and $Y_m$ being the mechanical loss angle and the Young's modulus of the $m-$th  layer, respectively.

According to eq (\ref{noise}),
increasing the beam-width $w$ and lowering the temperature $T$  result in a reduction of coating noise \cite{Harry}. 
Using wider (e.g., higher order Gauss-Laguerre) beams is another option, also being currently investigated \cite{Allocca}. 
Decreasing $T$ works for coating materials that does not exhibit mechanical loss peaks at the
(cryo) temperatures of interest \cite{Martin}. 
Current research is accordingly focused on
finding (synthetizing and optimizing)  {\em better} materials featuring low optical
absorption and scattering losses, low mechanical losses down to cryo temperatures, and high optical contrast (allowing fewer layers to
achieve a prescribed transmittance, resulting into thinner coatings and lower noise).
In this paper we focus on reducing the coating loss angle $\phi_c$ by optimizing the layer thicknesses.

\section{Coating optimization}

In this section we focus on the optimization of the coating structure sketched in Fig. 1. 
Let us use the suffixes S,L,H to identify substrate (S), low (L) and high (H) index material, respectively.
In the following we consider a coating consisting of $N_L$ layers  beginning with the
high refractive index materials\footnote{This is not restrictive, since the used optimization alghorithm is allowed to set the thickness of each and any layer to zero.} at the vacuum/coating interface .

As a consequence of the above assumptions 
\beq
\left \{
\begin{array}{ll}
n^{(m)}= n_H - \imath \kappa_H, \mbox{ $m$ odd}, & n^{(m)}= n_L - \imath \kappa_L, \mbox{ $m$ even};\\ 
\phi_m = \phi_H, \mbox{$m$ odd }, & \phi_m = \phi_L, \mbox{ $m$ even} ;\\ 
Y_m = Y_H, \mbox{ $m$ odd}, & Y_m = Y_L, \mbox{$m$ even }.
\end{array}
\right.
\eeq 
With this assumptions the coating  noise angle becomes:
\beq
\phi_c=  \eta_H \sum_{m \in J_o} d_{m}  + \eta_L \sum_{m \in J_e } d_{m},
\label{mirnoise}
\eeq
where $J_e=\{m \, \mbox{even integer}| 1 \le m \le N_L\}$ and $J_o=\{m \, \mbox{odd integer}|1 \le m \le N_L\}$.
Defining the normalized loss angle $\bar{\phi}_c= \phi_c /(\lambda_0 \eta_L)$ and introducing
the normalized physical length $z_m= d_m/\lambda_0$, 
where $\lambda_0$ is the free space wavelength, we have:
\beq
\bar{\phi}_c=    \sum_{m \in J_o} \gamma z_{m}  +  \sum_{m \in J_e} z_{m},
\label{normirnoise}
\eeq
where the noise ratio coefficient $\gamma = \eta_H/\eta_L$ can be explicitly written as:
\beq
\gamma = \frac{\phi_{H}}{\phi_{L}}\left( \frac{Y_{H}}{Y_s} + \frac{Y_s}{Y_{H}} \right) \left( \frac{Y_{L}}{Y_s} + \frac{Y_s}{Y_{L}} \right)^{-1}.
\label{gamma}
\eeq
In the case where the refractive index $n_L$ is the same as that $n_s$ of the substrate material (as in current GW detectors) 
$N_L$ is taken as an odd number\footnote{In fact choosing an even $N_L$  results in a configuration with the rightmost layer made of low refractive index material which
increases the noise without any effect  on the reflectivity.}.

\subsection{Constrained optimization formulation}
The optimization of the mirrors for GW detectors is a peculiar problem.
In standard mirror optimization design it is important to achieve high reflectance in a given frequency  and angular range. 
In the case of mirrors for GW detectors, the incidence is normal and the frequency range is very narrow (laser signal at $\lambda_0=1064 \, nm$ ) 
but it is mandatory to find a mirror setup that induces the minimal additional (thermal) noise on the detection channel.
Therefore it is important, in a suitable sense,  to reduce both the power  transmittance $\tau_c=1-|\Gamma_c|^2$ and the thermal noise loss angle $\bar{\phi}_c$.

As a consequence, a straightforward  formulation of the coating optimization problem for the design of low noise dielectric mirror can 
consist in searching for the thickness sequence that minimizes the thermal noise keeping the transmittance below a prescribed threshold value $\tau_0$.
This is a typical constrained optimization  problem \cite{COPTbook} that  in mathematical notation  can  be  written :
\begin{equation}
\begin{aligned}
& \underset{z_1,..., z_{N_L} \in \Omega}{\text{Minimize}}
& & \bar{\phi}_c \\
& \text{subject to}
& & \tau_c \leq \tau_0 \;
\end{aligned}
\label{optprob}
\end{equation}
where the constraint transmittance $\tau_0$ should be typically a few part per million (henceforth ppm).

Note that, in view of the transmittance constraint, problem (\ref{optprob}) is non-linear and non-convex. 
The search space $\Omega$ is defined by the inequalities $0 \le z_m \le 0.25/n_H$ for  odd $m$ and  $0  \le z_m \le 0.5/n_L$ for  even $m$.

An alternative way to formulate the optimization problem can
consist in searching for the thickness sequence that minimizes the transmittance keeping the thermal noise below a given threshold value:
\begin{equation}
\begin{aligned}
& \underset{z_1,..., z_{N_L} \in \Omega}{\text{Minimize}}
& & \tau_c  \\
& \text{subject to}
& & \bar{\phi}_c \leq \bar{\phi}_0 \; 
\end{aligned}
\label{optprob1}
\end{equation}
where $\bar{\phi}_0$ is a prescribed maximum allowed loss angle.
\subsection{Multiobjective optimization formulation}
The optimum coating design problem has been formulated in two alternative ways not necessarily
equivalent in eq.s (\ref{optprob}) and (\ref{optprob1}).
In this section,  we introduce a multiobjective optimization approach \cite{TutorialMOEA, MOEAbook} where we search for the thickness sequences minimizing simultaneously
the transmittance and the loss angle.
With  a non-standard mathematical notation we write:
\begin{equation}
\begin{aligned}
& \underset{z_1,..., z_{N_L} \in \Omega}{\text{Minimize}}
& & [ \bar{\phi}_c, \tau_c ] \\
\end{aligned}
\label{optprob2}
\end{equation}
Solving problem (\ref{optprob2}) in the framework of multiobjective optimization
means to find its tradeoff curve, also referred as the Pareto front or Pareto boundary,  in the $[\bar{\phi}_c, \tau_c ]$ plane. 
Each point belonging to the Pareto front corresponds to a sequence $z_m, \, m=1,...,N_L$ of layer normalized thicknesses.
 
In order to define the Pareto front of (\ref{optprob2}) 
 the concept of {\em dominance} \cite{TutorialMOEA}  has to be introduced, to define a suitable ordering rule in the $[\bar{\phi}_c, \tau_c ]$ plane. 
A physically feasible solution $\mathbf{A}$ in the space
$[ \bar{\phi}_c, \tau_c ]$ dominates another (different) physically feasible solution $\mathbf{B}$ 
if the coordinates of $\mathbf{A}$ are orderly less or equal to those of $\mathbf{B}$.
The set of physically feasible solutions, for which no physically feasible dominant solution exists,
is the Pareto front of the multiobjective optimization problem.
Let us note that the problem (\ref{optprob}) can be solved using the Pareto front of (\ref{optprob2}), 
by choosing the point on the Pareto front with transmittance component equal to $\tau_0$. Similar considerations can be done for
the problem (\ref{optprob1}), which  can be solved by taking the 
point on the Pareto front with noise component equal to $\bar{\phi}_0$. Furthermore, the properties and structure of the
Pareto front can give some hints on the relationship between problem formulations (\ref{optprob}) and
(\ref{optprob1}).

\section{Numerical solution of multiobjective optimization problem}

Many algorithms, based on different global multiobjective optimization tools, 
 are available in order to face the problem of Pareto front computation  for high dimensional  problems.
These algorithms generally use a suitable sampling method of the physical feasible configuration space, enabling the reconstruction of the Pareto front 
( e.g. NSGA-II, NSGA-III, $\epsilon$-MOEA etc. see \cite{MOEAbook}).

In this paper we perform a numerical exploration of the Pareto front (\ref{optprob2}) using 
a state of the art, public domain Multi Objective Evolutionary Algorithm (MOEA) named Borg MOEA \cite{borgMOEA},
that uses an evolutionary strategy appropriate for continuous variables. The algorithm 
is implemented as a package \cite{BlackBox} written in the Julia language \cite{Julia}.
The relevant literature and a simple description of the algorithm are reported in Appendix A.

Most multi objective algorithms use mutation, crossover  and selection operators, 
that do not change throughout the execution program.  
The Borg MOEA uses different operators from existing MOEAs and adopts them adaptively 
 on the basis of their success in the search.  
The evaluation of the progress, the adaptation of  the population size and the increasing of the archive with new solutions help the algorithm to continue the progress and prevents it 
from being trapped in some loop for the entire runtime.  
The goodness of numerical solutions given by the Borg MOEA  are compared with reference solutions.
We consider for comparison with the Borg MOEA optimized design, the coating structure which is currently used in Virgo/LIGO test masses \cite{pinard17}, 
consisting of alternating quarter wavelength layers of
SiO$_2$/ Ti-doped Ta$_2$O$_5$ deposited on a fused silica substrate. The physical parameters used in our simulations are reported in Table \ref{tab:partab}.

\begin{table}[t]
\centering
\begin{tabular}{|l| l|}
\hline \hline
Coating  & Substrate \\
H {\small (amorphous Ti-doped Ta$_2$O$_5$ )} &  \small (bulk crystalline SiO$_2$)\\  
L {\small (SiO$_2$)} &    \\  \hline
$n_H = 2.10$  & $n_s=1.45$ \\ \hline
$n_L = 1.45$  &  $Y_s = 72$ GPa  \\ \hline
$\kappa_H = 4.0 \times 10^{-8}$  & $\kappa_s=8.4 \times 10^{-11}$ \\ \hline
$\kappa_L =  8.4 \times 10^{-11}$  & $\phi_s = 7.00 \times 10^{-8}$   \\ \hline
$Y_H = 147$ GPa &   \\ \hline
$Y_L =  72$ GPa  &   \\  \hline
$\phi_H = 3.76 \times 10^{-4}$ & \\  \hline
$\phi_L = 5.00 \times 10^{-5}$ & \\  \hline
$\gamma = 9.5$ & \\  \hline
\hline
\end{tabular}
\caption{Physical parameters of coating and substrate materials used in simulations, we assume temperature $T=300 K$ and free space wavelength  $\lambda_0=1064 nm$.
}
\label{tab:partab}
\end{table}
\subsection{Pareto front convergence and structure}
It is well known that deterministic stopping criteria for global evolutionary optimization algorithms are unavailable.
Therefore, in order to investigate the convergence of the used algorithm, we computed the Pareto fronts with $N_L=11, \,15 , \, 19$  for 
increasing evolution times $T_s = 10^4 ,\, 2 \times 10^4 ,\, 5 \times 10^4 ,\, 10 \times 10^4 ,\, 20 \times 10^4$ sec.
The numerical estimated maximum absolute deviations (i.e. the uniform norm distance between Pareto curves at $T_s$ and $T_s/2$) are reported in Table \ref{tab:error}
 showing that the Pareto fronts do not change significantly for 
$T_s \ge 5 \times\, 10^4$  sec.
\begin{table}[t]
\centering
\begin{tabular}{|c|l|}
\hline \hline
Evolution Time $T_s$ [sec] & Absolute Error \\ \hline
$10 \times 10^3$ & $2.3 \times 10^{-3}$  \\ \hline
$20 \times 10^3$ & $1.0 \times 10^{-4}$ \\ \hline
$50 \times 10^3$ & $1.7 \times 10^{-5}$   \\ \hline
\hline
\end{tabular}
\caption{Estimated maximum absolute deviation (i.e. numerical estimation of the Pareto fronts distance in the uniform norm) 
between the Pareto fronts computed at time $T_s$ 
and the previous one computed at $T_s/2$, for the case $N_L=19$.}
\label{tab:error}
\end{table}
As a consequence, we took ({\em ad abundantiam}) $T_s \sim 10^5 $ sec in our simulations. 

In Fig. 2(a) we display Pareto fronts for the cases $N_L=11 \, (2) \, 19$, i.e. ranging from $11$ to $19$ with step $2$. 
It is seen that the Pareto fronts
 exhibit several bumps, whose number is equal to the number $N_D$ of high refractive index material layers. 
\begin{figure*}
\centering
\includegraphics[width=16 cm]{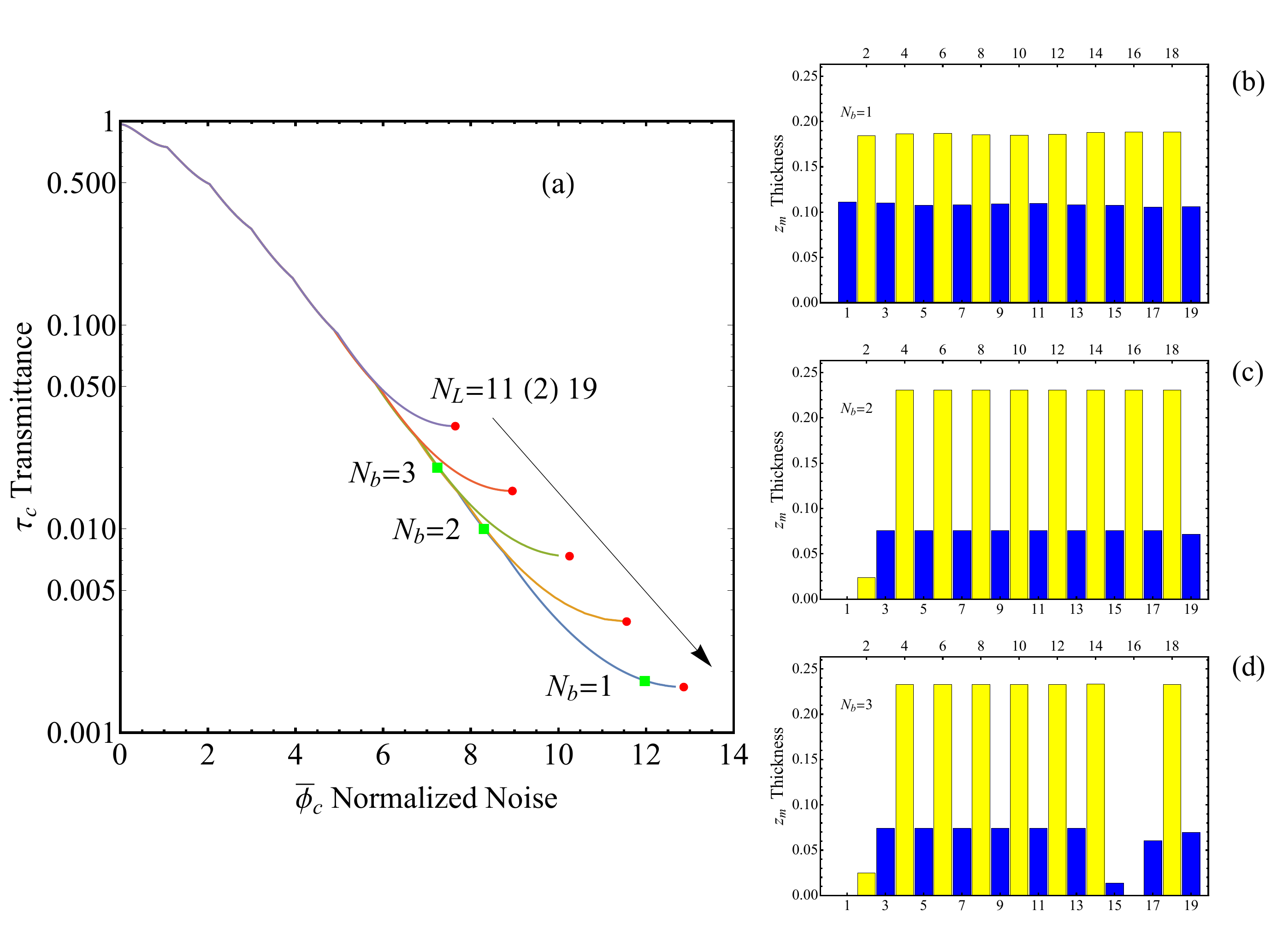}
\caption{ Left panel (a): Pareto fronts for the cases $N_L =11 \, (2) \, 19$. 
The bump-like structure of Pareto fronts is clearly shown in the high transmittance region of the curve.
For the case $N_L=19$ we indicate with $N_b=1,2,3$ the last three bumps. 
The red circular markers corresponds to the quarter wavelength design. 
In the right panel (b,c,d) we display, for the green square markers corresponding to the bumps $N_b=1,2,3$,  the thicknesses as a function of layer index $m$. 
The height of vertical blue (yellow) bars gives the  normalized layer thicknesses $z_m$ of  the high (low) refraction index material. }
\label{fig:2}
\end{figure*}

The bumps are less visible in the lower transmittance region.
In Fig. 2 (b),(c),(d) the sequences of layer thicknesses corresponding to the square green markers in \mbox{Fig. 2(a)} are displayed.
It can be noted  that the bump with lowest transmittance ($N_b=1$ in Fig. 2(b))
corresponds to a mirror configuration with all layer thicknesses different from zero. 
Moreover,  we note that the rightmost red point of the bump $N_b=1$ corresponds to the quarter wavelength design  at the operating wavelength $\lambda_0=1064 \, nm$ with $N_L=19$ .
This is in agreement with the well known quarter wavelength design property of 
minimizing transmittance in the cases of  multilayer reflectors made of negligibly absorbing materials \cite{carniglia80}. 
The next bump $N_b=2$ corresponds to a mirror design where the thickness of a single layer of high refractive index has been  {\it practically} set to zero  (see Fig. 2(c)). 
For the case $N_b=3$ illustrated in Fig. 2(d), the thickness of an additional layer of low index material is set to zero; this implies that the two nearby high refractive 
index layers merge toghether and can be considered as a single layer.
We found that the above behaviour can be generalized to all bumps,  i.e. the generic bump  $N_b=k$ fairly corresponds
 to  a multilayer structure with $N_D - k + 1$ high refractive index layers. 

This feature is a general characteristic of the Pareto front of dielectric mirror multiobjective optimization, which are, in summary, 
continuous, decreasing, and non-convex ({\em bumpy}) curves. The continuity of multiobjective tradeoff curves implies that the
problems (\ref{optprob}) and (\ref{optprob1}) are mathematically equivalent. 
In fact if we consider a generic point $[\bar{\phi}_0, \tau_0]$ on the Pareto front, we find that $\bar{\phi}_0$ is the solution of problem (\ref{optprob})
with constraint $\tau_0$, while  $\tau_0$ is the solution of problem  (\ref{optprob1}) with constraint $\bar{\phi}_0$.

\begin{figure}
\centering
\includegraphics[width=7.5 cm]{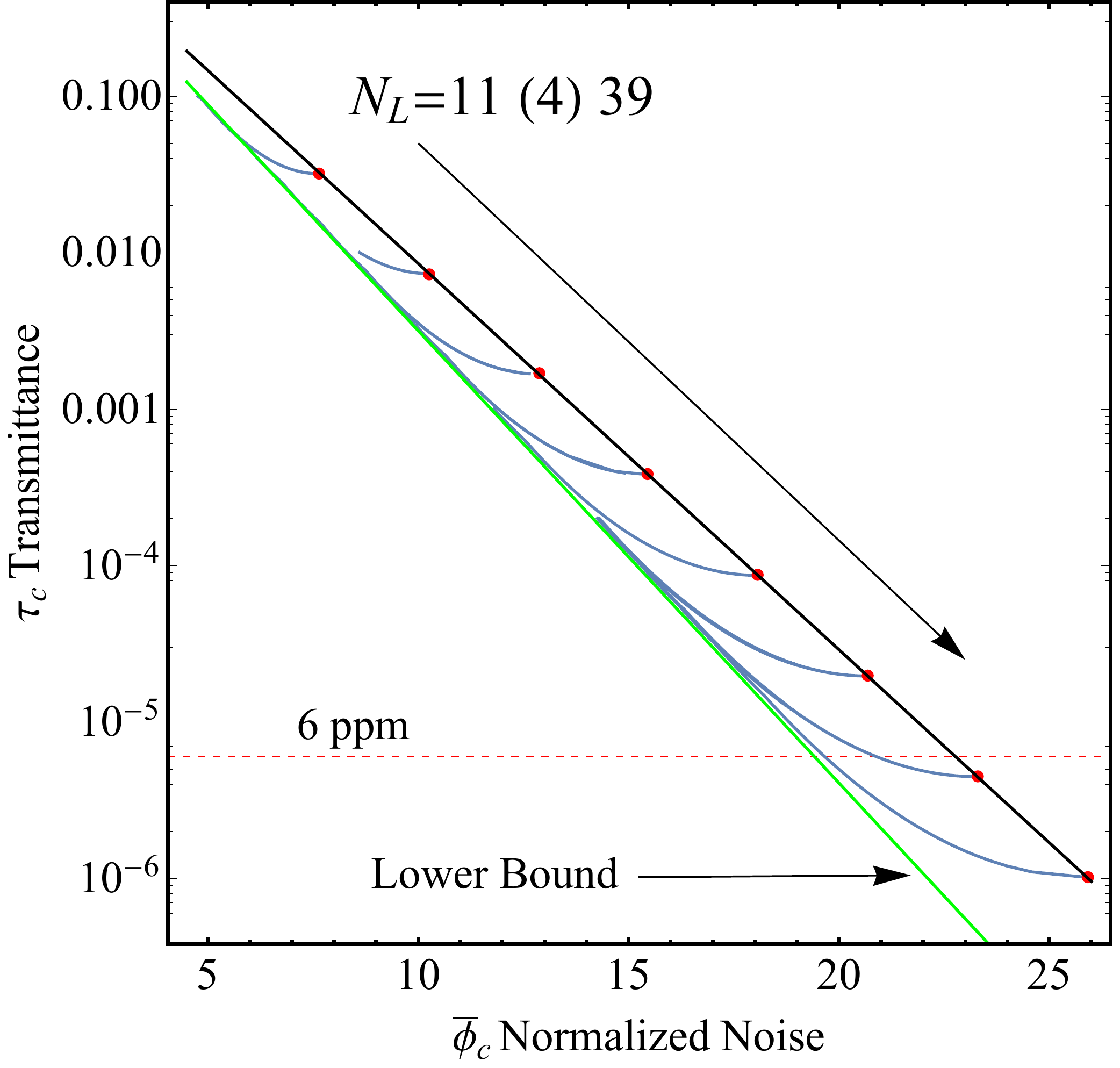}
\caption{Pareto fronts for $N_L=11$ to $N_L=39$ in step of $4$ in log-linear scale (blue curves). 
The simulation parameters are reported in Table \ref{tab:partab}.
Red circular markers refer to the quarter wavelength design. The dashed red horizontal grid line  at $\tau_c = 6$ ppm is the typical design target of
2nd generation (advanced) Virgo-LIGO detectors.
The black line corresponds to equation (\ref{trb}) computed with simulation parameters of Table \ref{tab:partab}, the green line
labeled "Lower Bound" is the same eq. (\ref{trb}) with an {\em effective} noise ratio coefficient $\gamma_e=7.9$.
}
\label{fig:3}
\end{figure}
In order to consider design configurations that exhibit performances of potential interest for present GW interferometer detectors, 
we performed simulations for coating multilayer up to $N_L=39$  with the  physical parameters reported in Table \ref{tab:partab}.
In Fig. 3 we display  the Pareto curves (blue lines) for $N_L$ ranging from $11$ to $39$ with step $4$. 
The red points in Fig. 3 are the quarter wavelength design which are located on a curve that can be written
 (under some approximations) as a straight line in the log-linear axes scale 
of the plane $[ \bar{\phi}_c, \tau_c]$.
The equation of the straight line is (see Appendix B for details):
\begin{eqnarray}
\displaystyle \log(\tau_c)= \log \left(\frac{4}{n_L} \right)-\frac{2 n_H}{\gamma n_L +n_H} \log \left(\frac{n_H}{n_L} \right) + \,\,\,\,\,\,\,\,\,\,\,\,  \nonumber \\
  \nonumber \\
\displaystyle \,\,\,\,\,\,\,\,\,\,\,\, - \bar{\phi}_c \frac{8 n_L n_H}{\gamma n_L +n_H} \log \left(\frac{n_H}{n_L} \right)\,. 
\label{trb}
\end{eqnarray}
The tradeoff curves asymptotically locate near to a straight line (green line in Fig. 3) in the plane $[ \bar{\phi}_c, \tau_c]$ with a log-linear axes scale.
The equation of this line turns out to be the same eq. (\ref{trb}) where the noise ratio coefficient is reduced to an {\em effective} 
value $\gamma_e=7.9$, computed by regression of Pareto front data.
We conjecture that this curve is a lower bound for all tradeoff curves in the region of low transmittance ($\tau_c \le 0.1$) above the Koppelmann  limit \cite{Koppelmann},
that is placed at $\tau_c \sim 10^{-7}$ for the parameters reported in Table \ref{tab:partab}. The Koppelmann limit is also the order of magnitude
of normalized energy absorption ({\em assorbance}) in the sought design in Fig. 2(b)(c)(d) and in the following.

The horizontal dashed red  line $\tau_c = 6 \, $ ppm shown in Fig. 3 corresponds to the target transmittance required for the design 
of dielectric mirrors used in GW detectors. The quarter wavelength coating with $N_L=35$ is the reference design because it
matches the target transmittance with the lowest normalized noise.

It is clear that each tradeoff curve with $N_L \ge 35$ contains layer configurations satisfying the constraint $\tau_c \le 6 \,$ ppm and
showing a reduced normalized noise $ \bar{\phi}_c$ with respect to the quarter wavelength reference design.

In figure 4(a)(b)(c) the layer thickness configurations obtained by the intersections between tradeoff curves with $N_L=39, 41, 43$ and the 
horizontal transmittance line $\tau_c = 6 \, $ ppm are displayed. 
These configurations give a reduction of the normalized noise
with the respect to the quarter wavelength reference design  of about $15.5\%, 15.9\%, 15.9 \%$ respectively.
 Note that the designs reported in Fig. 4(a) and 4(b) belong to the first  bump  of the tradeoff curves $N_L=\, 39, \, 41$ respectively,
while the configuration in Fig. 4(c) belongs to the second bump of the Pareto front $N_L=43$.

\begin{figure}
\centering
\includegraphics[width=8.5 cm]{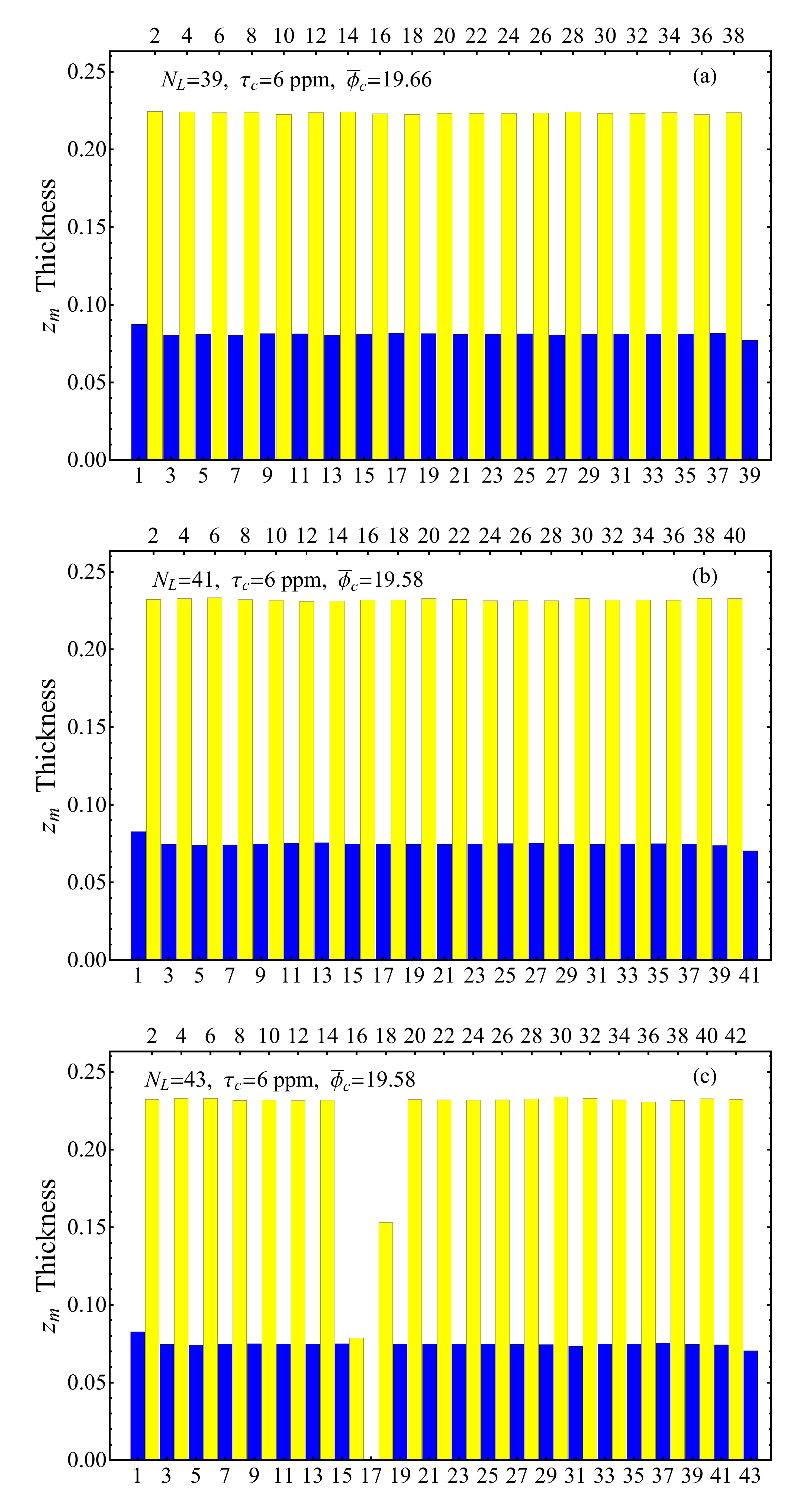}
\caption{From top (a) to bottom (c) the normalized layer thicknesses $z_m$  of 
the optimal multiobjective design for cases $N_L=39, \, 41, \, 43$.
The height of vertical blue bars gives the  normalized layer thicknesses $z_m$ of high refractive index material (odd $m$ values on the bottom axis) .
The height of vertical yellow  bars gives the normalized layer thicknesses $z_m$ of the low  refractive index material (even $m$ values on the top axis).
The design corresponds  to  the points obtained by the intersection of the Pareto boundary  with the horizontal line (dashed line in Fig. 3) drawn  
at the  prescribed transmittance level ($\tau_c=6 \,$  ppm).}
\label{fig:4}
\end{figure}

Moreover, let us note that the thickness sequences look like a truncated periodic configuration except for the first two layers and the last one.
The normalized physical length of the internal layers with high and low refractive index materials are $z_H \sim 0.081$ and $z_L \sim 0.22$, respectively, corresponding 
to normalized optical lengths $n_H \, z_H \sim 0.17$ and $n_L \,z_L \sim 0.32$, whose sum is less than $0.5$.

\begin{figure}
\centering
\includegraphics[width=9 cm]{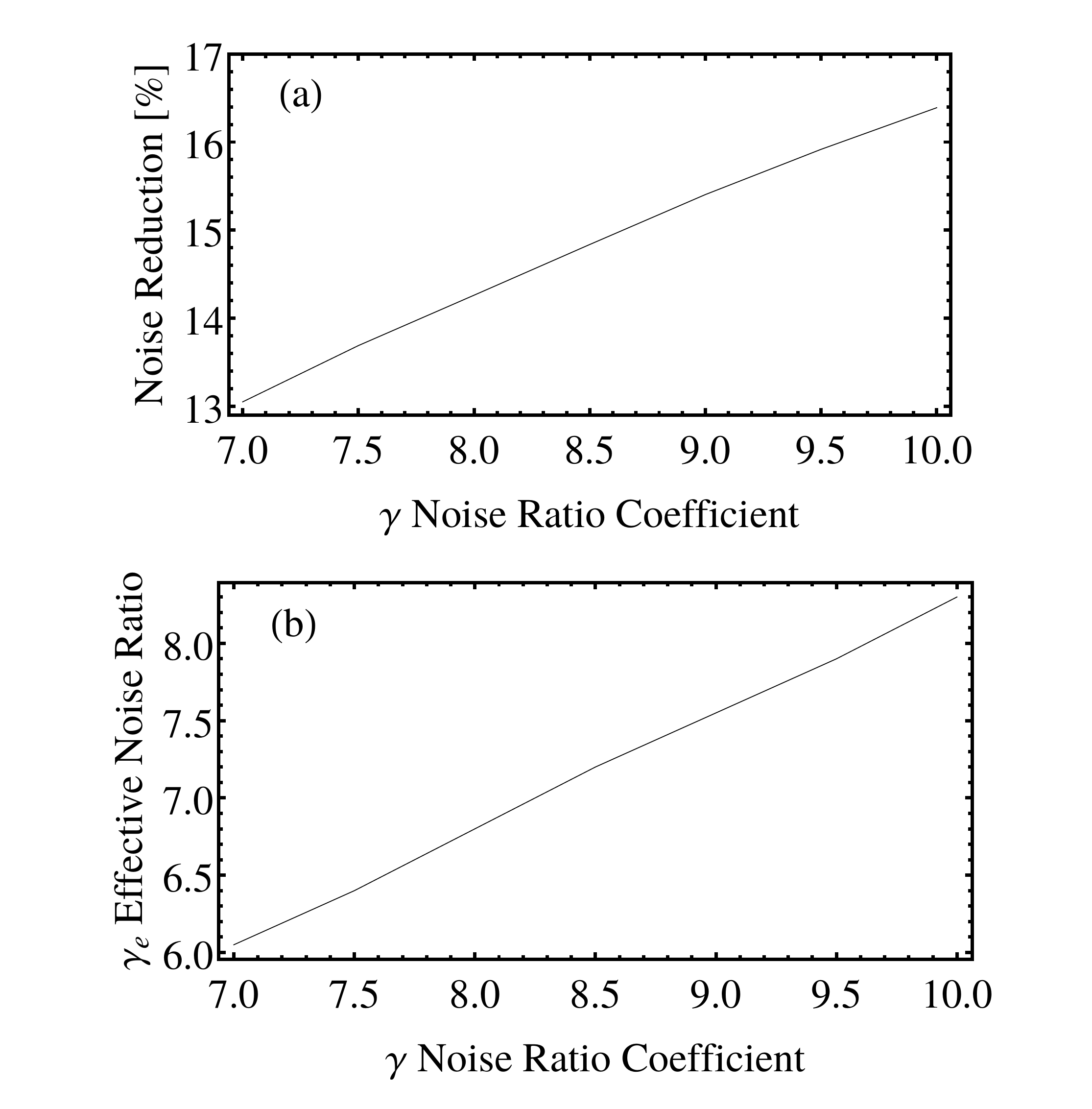}
\caption{  (a) The percent loss angle reduction of the optimizated coatings with respect to the reference quarter wavelength design. 
(b) The {\em effective} noise ratio coefficient $\gamma_e$  as a function of $\gamma$, computed by fitting eq. (\ref{trb}) to the Pareto front.}
\label{fig:5}
\end{figure}

In view of the uncertainty in the measurements of the specific loss angle coefficient,
we have performed all the above simulations considering the same refractive indexes $n_H$, $n_L$ and $n_s$  (see Table \ref{tab:partab}) and changing the noise ratio coefficient  
in the  range $\gamma \in [7,\, 10]$.
All of the above findings are confirmed.  We report in Fig. 5 a summary of the relevant results.
 In particular, the noise reduction of the optimized coating structure
with respect to the reference  quarter wavelength design  and the effective $\gamma_e$ noise ratio coefficient 
are reported as  functions of $\gamma$ in  Fig. 5(a) and 5(b), respectively. 
We have shown that both the noise reduction ({\it gain}) and the effective noise ratio $\gamma_e$ are almost linear, increasing  functions of the noise ratio coefficient $\gamma$.

The above parametric exploration has been done exploiting the possibility in reducing  the computational burden 
for the calculation of the limiting curve of Pareto fronts. Indeed the bound given by eq. (\ref{trb}) can be evaluated only 
for two value of $N_L$ in a moderately low $\tau_c$ region, 
and than extrapolated via  mono-parametric regression to the whole low transmittance region above the Koppelman limit.

\begin{table}[t]
\centering
\begin{tabular}{|l|l|l|l|}
\hline
\hline
\tiny \, &     \,   \tiny      &  \,  \tiny      &   \, \tiny             \\
Parameters \, &     \,   $\bar{\phi}_c$      &  \,  $\bar{\phi}_c$      &   \, $\bar{\phi}_c$             \\
   &  \small Periodic &  \small Tweaked & \small Borg MOEA \\ \hline \hline
\begin{tabular}[c]{@{}l@{}}$N_L=19$ \\ $\tau_c=6\, \times 10^{-3}$\end{tabular} &  9.1214  &   9.1054  &  9.1054  \\ \hline
\begin{tabular}[c]{@{}l@{}}$N_L=27$ \\ $\tau_c=10^{-4} $\end{tabular}    & 16.5918  & 16.5879 & 16.5879  \\ \hline
\begin{tabular}[c]{@{}l@{}}$N_L=41$ \\ $\tau_c=6\, \times 10^{-6}$\end{tabular} & 19.5968   & 19.5809 & 19.5809  \\ \hline \hline
\end{tabular}
\caption{Normalized noise $\bar{\phi}_c$ of the optimum design. The target transmittance $\tau_c$ and the number of layers $N_L$ are reported in the first column.
For these parameters, the minimal loss angles, obtained with different optimization procedures are displayed in the other columns.}
\label{tab:tweaked}
\end{table}

\subsection{The tweaked periodic solution and the mixture checks}

We have found that the thickness sequences look like a truncated periodic configuration with the exception of the two  topmost layers and the last one.
This result corroborates the working hypothesis used in previous papers \cite{Spie, Villar10}.
As a further check, we have implemented a code to face the tweaked periodic truncated design problem where the search space $\Omega$ is reduced 
to a five-dimensional space  by setting
$z_1=x_1, \, z_2=x_2$, and  $z_{N_L}=x_5$   for the first two layers and the last one in the dielectric mirror.
In the remaining layers for all $m$ odd in the range $N_L > m > 1$ we set $z_m = x_3 $, and  for all $m$ even  in the range  $N_L > m > 2$
we set $z_m = x_4$.
In Table \ref{tab:tweaked} we report  the tweaked and the complete multiobjective solutions. We note
that  in terms of noise reduction an agreement on the first four decimal figures is found.
We have also considered the periodic truncated design with a two-dimensional search space $\Omega$. 
The results shown in Table \ref{tab:tweaked}  evidentiate a deviation 
of the periodic truncated design with respect to the full evolutionary solution.
This difference in the tradeoff curve increases in the low transmittance region.

A final check has been carried out in order to verify whether it is convenient to substitute the high refractive
index material (noisier material) with a layered mixture of this material and the low index one.
To this end we expand the search space including for the high refractive layers (modeled like a mixture with layered inclusion) 
an additional variable $\xi_i$ which account
for the percent of high refractive index material present in the layered mixture.
In the case of layered inclusion the mixture refractive index  can be modeled following \cite{Sihvola, mixture}  and is $\sqrt{n_H^2 \xi_{i} + n_L^2 (1-\xi_{i})}$ , while 
the normalized thermal noise ratio becomes $\gamma \xi_{i} + (1-\xi_{i})$.
The multiobjective simulations performed in the expanded $N_L+N_D$-dimensional space $\Omega$, 
 using the same parameters of Table \ref{tab:partab} and for different values of $N_D$, 
shown that the Pareto tradeoff curve does not change because the
algorithm automatically finds that $\forall i: \,\,\, \xi_i \sim 1$.
This result implies that mirror design made of {\em true} (i.e. with $\xi_i < 1$) layered mixture solutions are located above the binary coating Pareto fronts.

The failure of this check definitively gives an answer on the optimality of mirror 
design with subwavelength layered  inclusions  in  the case of  coatings made of two materials,  
and also establish that it is not convenient to arbitrarily increase $N_L$,
keeping fixed $\tau_c$,  to search the optimal solutions.

\section{Conclusions}
In this article we faced the problem of optimizing the design of binary (two-materials) coatings for GW detectors 
by formulating and solving a multiobjective optimization problem. 
This approach  allowed us to  minimize  both the transmittance and the thermal noise.
While the values of the refractive indexes and the number of the layers are given, 
no {\em a priori } hypothesis on the thickness of the layers was made in the multiobjective optimization problem,
tackled using a global optimization method (Borg MOEA)
in a search space with a dimension equal to the number of layers.

We have shown by extensive numerical simulation based on Borg MOEA algorithm  the existence of a Pareto tradeoff boundary which is a continuous, 
decreasing, and non-convex (bump-like) curve. 
In particular, continuity implies that the multiobjective approach (\ref{optprob2}) 
is equivalent to  the  constrained  single-objective optimization problems (\ref{optprob}) and (\ref{optprob1}).
A solution (i.e. a mirror design) on the Pareto front corresponds to a sequence of layer thicknesses.

The thicknesses sequences on the Pareto boundary look like truncated periodic configurations except for the
first two layers and the last one.
In view of this result (see also \cite{Spie}), we faced the optimization problem in a search space with a reduced dimension.
We have considered the adapted periodic (five-dimensional search space) and  the truncated periodic (two-dimensional search space) sequence.
The performance of the tweaked (adapted) periodic sequences are comparable to the performance of the sequences obtained by the solution of the full multiobjective problem. 
The periodic design is outperfomed by the multiobjective design especially in the low trasmissivity region.

We have also shown that the Pareto fronts are bounded from below by an
exponential curve in the transmittance-noise plane (\ref{trb}). 
This curve has the same expression of the approximate relation between transmittance and noise for the quarter  wavelength 
design, except for the noise ratio coefficient which assumes a reduced {\em effective} value.
A possible application of eq. (\ref{trb}), consisting in the reduction of the computational burden 
for the calculation of the Pareto fronts, has been implemented for a parametric exploration in a suitable $\gamma$ range.

The noise reduction ($\sim 16\%$ for realisitc cases) has been shown to be (more or less) a linear, increasing 
function of the noise ratio coefficient $\gamma$. 

There is no possibility to ameliorate these performances by using instead of high refractive index material 
a  mixture  made of layered inclusions of low refractive index materal, with subwavelength thicknesses, placed inside the high refractive index material.

We are confident that the method used in this paper can be generalized to the analysis of the coating design
of dielectric mirror with three or more different materials, to be discussed in a future paper.

\section*{Acknowledgments}
\vspace{0.0cm}

This work has been partially supported by INFN through the projects Virgo and Virgo$-$ET.
M. Principe acknowledges the L'Oreal-UNESCO For Women in Science Program for supporting her in this work.
The authors are grateful for the discussion and suggestions received from the Virgo Coating R\&D Group and the Optics Working Group of the LIGO
Scientific Collaboration.

\vspace{1.2cm}
\section*{Appendix A - evolutionary algorithm in a nutshell}
\vspace{0.cm}

The evolutionary algorithms are smart versions of random search derived from the genetic evolution theory and
implemented on digital computers. They are well established classical tools for global optimization techniques
 \cite{Holland}.

To solve optimization problems with an evolutionary algorithm the individuals of a population  are
associated, by the encoding procedure, to a physical solution of a given problem (in our case the layer thicknesses sequence), 
the selection probability is proportional to the
quality of the represented solution, i.e. to the fitness function to be optimized.

The population  then undergoes selection crossover and mutation (like in natural genetics evolution), producing new children and updating the population.
The process is repeated over various generations until a suitable termination criteria is reached. 
Each individual, encoding a candidate solution, is assigned a fitness value 
, based on its objective function value, and the fitter individuals are given a higher chance to mate and yield more {\em fitter} 
individuals. 

The multi objective version of evolutionary algorithm  follows the same schema (see  \cite{TutorialMOEA, MOEAbook} for details), 
with the addition of a suitable strategy of Pareto front extraction.


\vspace{0.1cm}
An evolutionary heuristic follows this basic scheme:

\begin{algorithm}
 \KwResult{Pareto front of population}
 initialize random population $P(0)$, at the iteration $t=0$ \;
 find fitness of initial population\;
 extraction of Pareto front elements\;
 \While{termination criteria is not reached}{
  parent selection\;
  crossover of the parent\;
  mutation\;
  decode and fitness calculation\;
  survivor selection and update population $P(t+1)$\;
  extraction of Pareto front elements from $P(t+1)$\;
  increment $t$\;
 }
 \Return Pareto fronts\;
 \caption{Multi-Ojective evolutionary algorithm}
\end{algorithm}
\vspace{0.7cm}
\[
\,
\]
We address the reader to the cited literature \cite{borgMOEA,BlackBox} for a detailed description of the used algorithm.

\section*{Appendix B - transmittance vs thermal noise for the quarter wavelength design}

In this appendix we compute the transmittance  as a function of thermal noise for a quarter wavelength coating.
The reflectivity of a quarter wavelength multilayer made of $N_L=2 N_D-1$ alternating layers with high and low refractive indexes $n_H$ and $n_L$,
respectively,  placed on a substrate with refractive index $n_s=n_L$ is:
\beq
|\Gamma_c|^2=\frac{\left ( 1- n_L \left(\frac{n_H}{n_L}\right)^{2N_D} \right)^2}{\left ( 1+ n_L \left(\frac{n_H}{n_L}\right)^{2N_D} \right)^2}
\label{Am1}
\eeq
where $n_H/n_L \ge 1$. In view of eq. (\ref{Am1})  the transmittance $\tau_c=1-|\Gamma_c|^2$ reads:
\beq
\tau_c=
\frac{4 n_L \left(\frac{n_H}{n_L}\right)^{2N_D}}{\left ( 1+ n_L \left(\frac{n_H}{n_L}\right)^{2N_D} \right)^2}
\eeq
that can be written as
\beq
\tau_c=\frac{\frac{4}{n_L} \left(\frac{n_H}{n_L} \right)^{-2N_D}}{\left(1 + \frac{1}{n_L} \left(\frac{n_L}{n_H} \right)^{2N_D}\right)^2}
\label{A0}
\,\,\,.
\eeq
For  $\frac{1}{n_L} \left(\frac{n_L}{n_H} \right)^{2N_D} \ll 1$, by applying the $\log$ on both side of the
 equation (\ref{A0}) we get
\beq
\log(\tau_c) \sim \log \left(\frac{4}{n_L} \right) -2 N_D \log \left(\frac{n_H}{n_L} \right)
\,\,\,.
\label{A1}
\eeq
On the other hand, specializing eq. (\ref{normirnoise}) for the quarter wavelength design,  the normalized loss angle $\bar{\phi}_c$
can be written as a function of $N_D$
\beq
\bar{\phi}_c = \frac{\gamma N_D}{4 n_H} +  \frac{N_D-1}{4 n_L} = ( \frac{\gamma}{4 n_H} +  \frac{1}{4 n_L}) N_D - \frac{1}{4 n_L}
\label{A1.5}
\,\,\,. 
\eeq
Solving (\ref{A1.5}) for $N_D$ and plugging into (\ref{A1}) we have:
\[
\log(\tau_c)\sim \log \left(\frac{4}{n_L} \right)-\frac{2 n_H}{\gamma n_L +n_H} \log \left(\frac{n_H}{n_L} \right)+
\] 
\beq
\,\,\,\,\, -\bar{\phi}_c \frac{8 n_L n_H}{\gamma n_L +n_H} \log \left(\frac{n_H}{n_L} \right)\,. 
\label{transmbound}
\eeq

\end{document}